\documentclass[aps,prl,twocolumn,groupedaddress,floatfix,showpacs,showkeys,amsmath,amssymb,10pt]{revtex4-1}
\usepackage{graphicx}
\usepackage{bm}
\begin{document}
\title{Electronic properties of 8-{\it Pmmn} borophene}

\author{Alejandro Lopez-Bezanilla$^{1}$}
\email[]{alejandrolb@gmail.com}
\author{Peter B. Littlewood$^{1},^{2}$}
\affiliation{$^{1}$Argonne National Laboratory, 9700 S. Cass Avenue, Lemont, Illinois, 60439, United States}
\affiliation{$^{2}$James Franck Institute, University of Chicago, Chicago, Illinois 60637, United States}

\begin{abstract}
First-principles calculations on monolayer 8-{\it Pmmn} borophene are reported to reveal unprecedented electronic properties in a two-dimensional material. 
Based on a Born effective charge analysis, 8-{\it Pmmn} borophene is the first single-element based monolayered material exhibiting two sublattices with substantial ionic features.
The observed Dirac cones are actually formed by the p$_z$ orbitals of one of the inequivalent sublattices composed of uniquely four atoms, yielding an underlying hexagonal network topologically equivalent to distorted graphene. A significant physical outcome of this effect includes the possibility of converting metallic 8-{\it Pmmn} borophene into an indirect band gap semiconductor by means of external shear stress. The stability of the strained structures are supported by a phonon frequency analysis. The Dirac cones are sensitive to the formation of vacancies only in the inequivalent sublattice electronically active at the Fermi level. 
\end{abstract}
\maketitle

\section{Introduction}
  
The discovery of new monolayered materials is rapidly occurring due to the remarkable progress in synthesis and characterization of compounds at the nano-scale. 
Simultaneously, {\it in silico} methods are allowing to speed the pace of compound discovery while reducing the need for demanding experimental work. 
In contrast to the reduced number of experimental works on boron allotropes, numerous computational studies have predicted a plethora of two-dimensional (2D) boron based layers and clusters with various geometries and symmetries\cite{PhysRevLett.112.085502,PhysRevLett.99.115501}. Buckled deformations and vacancies were proposed to enhance the stability of B clusters, and noble-metal substrates to provide weakly interacting surface for extended layer growth. 
Stable phases of boron were portended through a series of computational studies based on evolutionary algorithms, with the so-called $\alpha$-sheet as the most stable structure \cite{PhysRevLett.112.085502}. Several 2D boron structures have been predicted to exhibit superconductivity with the critical temperature
above that of liquid hydrogen \cite{PhysRevB.93.014502}.

Continuing to improve the experimental techniques has led to the experimental fabrication of borophene \cite{Mannix1513}, a single layer of boron atoms which structure had been described via computer simulation but comprised numerous experimental challenges. 
Mannix et al.\cite{Mannix1513} reported the synthesis of extended clusters of boron in ultrahigh vacuum conditions over an inert Ag substrate. 
Scanning tunneling microscopy characterization of the one-atom-thick 2D sheets revealed a hexagonal arrangement of boron atoms with an extra atom in the middle of each hexagon in a buckled geometry to accommodate the greatest number of B atoms on the adsorption sites. As a result of the several electronic states crossing the Fermi level, this structure exhibits a robust metallic behavior \cite{C6TC00115G}. In a different experimental realization Feng et al. \cite{FengB} reported the synthesis of triangular lattices but in a flat geometry and with a periodic arrangement of atom vacancies.

However, in the absence of the substrate, thermal vibrations are expected to affect the stability of the suspended sheet and induce lattice distortions. The trend of boron to form buckled surfaces, providing the layer with a finite thickness, points towards structured layers as the most stable all-B extended sheets. The deposition of B on less interacting $\pi$-conjugated substrates such as graphene may stabilize the 8-{\it Pmmn} structure (8 stands for the number of atoms in the unit cell belonging to the space group {\it Pmmn}).
{\it Ab initio} studies reported that this B allotrope may exhibit distorted Dirac cones, massless electrons, and dynamical stability, but in a rhombohedral symmetry. Its formation energy per atom is below the 2D $\alpha$-structure and hundreds of meV above the bulk $\alpha$-structure \cite{PhysRevLett.112.085502}. It is therefore worthwhile to analyze in detail the electronic features of the 8-{\it Pmmn} boron allotrope, which further exhibits enhanced stability in the free-standing form. 

\begin{figure}[htp]
 \centering
  \includegraphics[width=0.4 \textwidth]{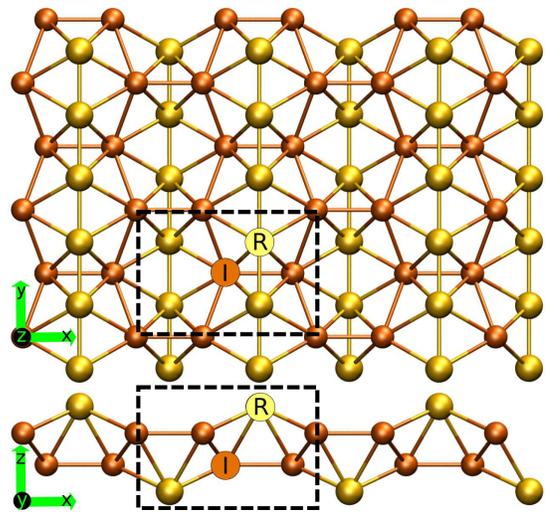}
 \caption{Top and side view of 8-{\it Pmmn} borophene. The color-scheme distinguishes the two types of nonequivalent B atoms, namely the atoms along the nano-structure ridges (B$_R$), and the atoms bonded to the them in inner positions (B$_I$). Dashed rectangles delimit the unit cell.}
 \label{fig1}
\end{figure}

In this paper we provide a systematic study of the electronic properties of 8-{\it Pmmn} borophene, unveiling the origin of the Dirac cones and the possibility to open a band gap under external physical stimuli. A simple argument that relates the Dirac cones to an underlying honeycomb-like lattice is provided, in addition to a description of the ionic character of the nonequivalent B atoms in the lattice.

\section{Methods}

Density functional theory based calculations were performed within the Perdew-Burke-Ernzerhof (PBE) generalized gradient approximation (GGA) functional for the exchange-correlation, and the projector-augmented-wave method as implemented in VASP\cite{PhysRevB.48.13115,PhysRevB.54.11169,PhysRevB.59.1758}. The electronic wave functions were computed with plane waves up to a kinetic-energy cutoff of 500 eV. The integration in the k-space was performed using a 40 $\times$ 50 $\times$ 1 Monkhorst-Pack k-point mesh centered at $\Gamma$-point. Atomic coordinates and lattice constants were fully relaxed until the residual forces were smaller than 10$^{-3}$ eV/\AA. The hybrid HSE06 functional as implemented in VASP was additionally utilized to confirm the formation and shape of Dirac cones. The force-constant method and the PHONOPY package \cite{phonopy} were employed for computing phonon spectra, and the dynamical matrices were computed using the finite differences method in large supercells.  

The geometry optimizations and electronic structure calculations of supercells containing up to 280 atoms with a single vacancy were obtained with the SIESTA code \cite{PhysRevB.53.R10441,0953-8984-14-11-302}. A double-$\zeta$ polarized basis set within the GGA approach for the exchange-correlation functional was used. Atomic positions were relaxed with a force tolerance of 0.01 eV/\AA. The integration over the Brillouin zone was performed using a Monkhorst-Pack sampling of 5$\times$5$\times$1 k-points for 2.26 nm $\times$ 2.29 nm large supercells containing a vacant site.

\section{Results and discussion}

\subsection{Effective charges}

According to the symmetry of 8-{\it Pmmn} borophene the unit cell contains two types of nonequivalent B atoms, namely ridge atoms (B$_R$) and inner atoms (B$_I$). Figure \ref{fig1} shows a model where B$_R$ and B$_I$ atoms are labeled and colored distinctively. The orthorhombic unit cell is defined by a 4.51 \AA\ $ \times$ 3.25 \AA\ rectangle. The vertical relative distance between two B$_R$ atoms is 2.20 \AA. The different local environment of the B$_R$ and B$_I$ atoms suggests that they could exhibit distinct physical and chemical properties. 
This is qualitatively confirmed by a Mulliken population analysis of the electronic distribution. According to the results obtained with the SIESTA code a charge of 3.26 $\mid$e$\mid$ for each B$_R$ atom and of 2.74 $\mid$e$\mid$ for each B$_I$ atom ($\mid$e$\mid$ is the Fundamental Charge Unit). Therefore, half of an electron migrates from the inner to the ridge boron atoms, rendering 8-{\it Pmmn} borophene as a single-element 2D material that, being covalent, exhibits an ionic character.

\begin{table}
\begin{ruledtabular}
\begin{tabular}{|c c c c c c c c|}
\hline 
 [$|e|$]& \multicolumn{1}{c}{$Z^*_{xx}$} & \multicolumn{1}{c}{$Z^*_{yy}$} & \multicolumn{1}{c}{$Z^*_{zz}$}& \multicolumn{1}{c}{$Z^*_{yz}$}& \multicolumn{1}{c}{$Z^*_{zy}$} & \multicolumn{1}{c}{$Z^*_{zx}$}& \multicolumn{1}{c}{$Z^*_{xz}$} \\ 
\hline
 B$_I$& -1.33 & -0.78 & -0.06 & 0.0 & 0.0			 & $\pm$1.27& $\pm$0.03			\\
 B$_R$& 1.33  & 0.78  &  0.06 & $\pm$0.10& $\pm$1.41& 0.0 			&0.0 \\
\hline 
\end{tabular}
\end{ruledtabular}
\caption{The dynamical (Born) effective charges (in units of the Fundamental Charge Unit $|e|$) of the B$_I$ and B$_R$ atoms in 8-{\it Pmmn} borophene. The acoustic sum rule $\sum_\alpha Z^*_\alpha=0$ is satisfied for each matrix component $\alpha$ of the Born charge matrix.}
\label{table}
\end{table}

To further confirm this trend and measure the deviation from a purely covalent bond, we use density-functional perturbation theory to calculate the Born (dynamical) effective charge tensor, Z$^*_B$, of each B atom. The Born charge matrix represents the electrical polarization $\vec P$ produced by a displacement $\vec u$: $P_\alpha = |e|Z_{\alpha\beta}u_{\beta}$.  According to the diagonal terms of the tensor (Table \ref{table}), B$_R$ and B$_I$ exhibit a considerable anisotropy along the three axis with equal but of opposite sign effective charges. This implies a substantial IR optical activity of the corresponding optical phonon modes. Both the size (larger than the Mulliken charges) and the symmetry (the existence of large off-diagonal terms) cannot be explained in a rigid ion picture. Clearly there is substantial redistribution of the electronic bond charge in response to nuclear motion - as seen in, for example, the narrow gap elemental materials Se and Te \cite{Littlewood1984}. The two mirror planes in the unit cell rule out any permanent dipole in the co-planar direction, and in the z-direction there is a symmetry operation of a mirror combined with a translation that will rule out the third polar component, pointing out to the existence of a quadrupole.

\begin{figure*}[htp]
 \centering
  \includegraphics[width=0.99 \textwidth]{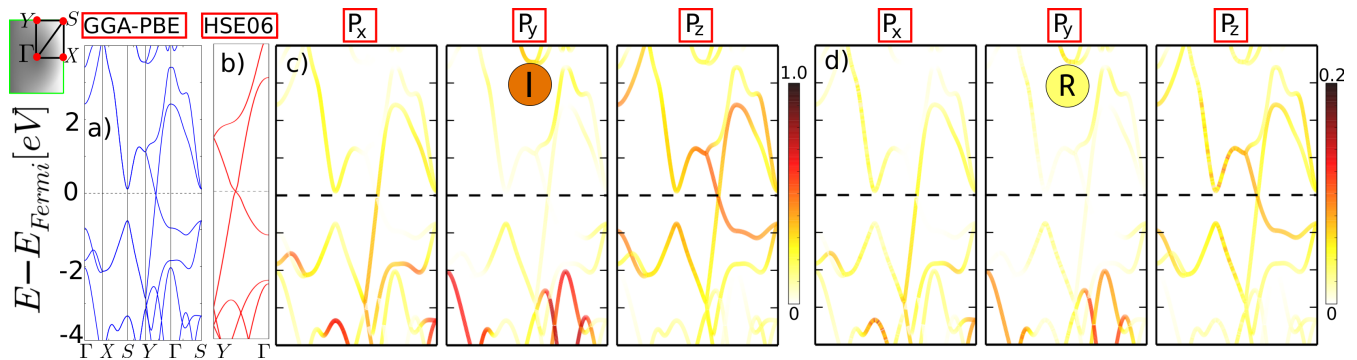}
 \caption{a) displays the GGA-PBE computed electronic band diagram of 8-{\it Pmmn} borophene. In b), the Dirac cone in the $Y \rightarrow \Gamma$ direction in reciprocal space is computed with the HSE06 functional. The contribution of B$_R$ and B$_I$ atoms to the formation of the Dirac cone within GGA-PBE approach is unveiled in the color-resolved band diagrams for the B$_I$ atoms, c), and the B$_R$ atoms, d). For the sake of clarity the color normalization in c) is set to 1.0, while in d) is set to 0.2. Therefore the Dirac cone is mostly formed by the B$_I$ atoms.
 }
 \label{fig2}
\end{figure*}

\subsection{Electronic structure}

The non-equivalence of the B$_R$ and B$_I$ atoms is also manifested in the contribution of each type of atoms to the electronic states of the material. Figure \ref{fig2}a displays the electronic band diagram of 8-{\it Pmmn} borophene. Two dispersive bands touch each other in the high-symmetry line connecting the $Y$ and $\Gamma$ points in reciprocal space. In the Brillouin zone there are actually two Dirac cones related by symmetry operations, of which only one is shown. Within the GGA-PBE formalism, a Dirac cone with linear dispersion in the Dirac point is observed. The HSE06 functional yields a slightly different result (Figure \ref{fig2}b), namely electronic states with an energy extension that surpasses the former and, more importantly, a Dirac cone with a less pronounced dispersion in the vicinity of the Dirac point. Two turning points close to the Dirac point reduce the slope of the bands and subsequently the effective mass of the charge carriers. For a sufficiently dense k-mesh in the reciprocal space (48 $\times$ 60 $\times$ 1), no hole pocket is observed in the band diagram obtained with the HSE06 functional, contrary to Ref. \cite{PhysRevLett.112.085502} where a low-dense k-mesh was employed.

To elucidate the contribution of each atom to the formation of the electronic bands in the vicinity of the Fermi level, we resort to the color-weighted representations of the p orbitals of both types of atoms, as shown in Figure \ref{fig2}c and d for B$_I$ and B$_R$ atoms respectively.
(Note that, for the sake of clarity, the color normalization is different for each type of atom, from 0 to 1 for B$_I$ and from 0 to 0.2 for B$_R$).
The contribution of the s-orbitals is negligible and is not included in the Figure.

By observing the band diagrams it is easy to identify the p$_z$ orbital of the B$_I$ atoms as the main one responsible of the formation of the Dirac cones, with a much lesser contribution of the p$_x$ orbital and only to one of the bands. On the contrary, the B$_R$ atoms are ineffective in the formation of the electronic states in the vicinity of the Fermi level, and their participation is residual. This is in contradiction and corrects the analysis of the first study on 8-{\it Pmmn} borophene (Ref. \cite{PhysRevLett.112.085502}) where it was stated that the in-plane p$_x$ orbitals from the buckled boron chains, and out-of-plane states p$_z$ orbitals from the buckled irregular boron hexagons are responsible for the emergence of Dirac cone.
 
\begin{figure}[htp]
 \centering
  \includegraphics[width=0.49 \textwidth]{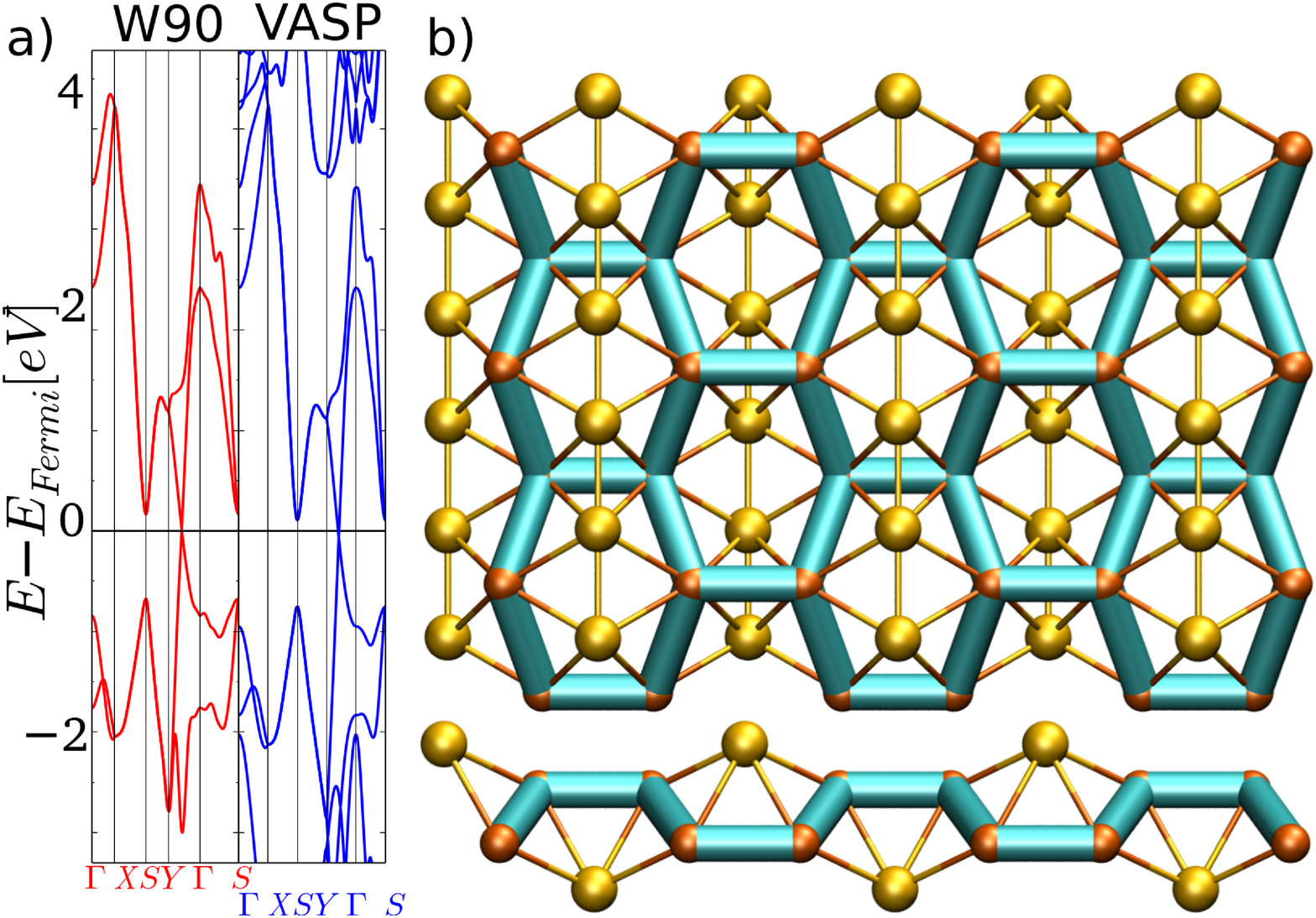}
 \caption{a) Comparison between the energy band diagram generated with the Wannier90 interpolation method by projecting Bloch wave functions onto the p$_z$ orbitals of the B$_I$ atoms (left, red lines), and the DFT computed band diagram (right, blue lines). b) shows the effective hexagonal network of B atoms generating the Dirac cone superimposed to the 8-{\it Pmmn} borophene structure.}
 \label{fig3}
\end{figure}

To further corroborate this explanation on the origin of the Dirac cones, we resort to maximally localized Wannier functions (MLWFs)\cite{PhysRevB.56.12847} to derive a model Hamiltonian of the system. MLWFs were obtained using the Wannier90 code\cite{Mostofi2008685} from the first-principles calculation of the ground state. We aim at reproducing the two valence and two conduction bands in the vicinity of the Fermi level within a four-band model. By projecting onto the p$_z$ orbitals of the B$_I$ atoms and minimizing the MLWF spread the band structure obtained using the Wannier90 interpolation method (Figure \ref{fig3}a) is in very good agreement with the first-principles band structure (Figure \ref{fig2}a). Figure \ref{fig3}b shows a schematic representation of the resulting effective honeycomb-like network of B$_I$ atoms. The Dirac cones of 8-{\it Pmmn} borophene emerge from this hexagonal lattice which is topologically equivalent to uniaxially strained graphene. 

Although 8-{\it Pmmn} borophene  has a finite thickness with nonequivalent atoms within a rectangular unit cell, the electronic properties of this B allotrope do not differ from other single-element 2D materials, since only the atoms in a hexagonal network (graphene) that are in a buckled geometry (silicene) lead to the formation of the Dirac cones. What makes this material particularly interesting is the possibility of distorting the Dirac cone and open an electronic band gap.

\subsection{Shear stress}

\begin{figure}[htp]
 \centering
  \includegraphics[width=0.49 \textwidth]{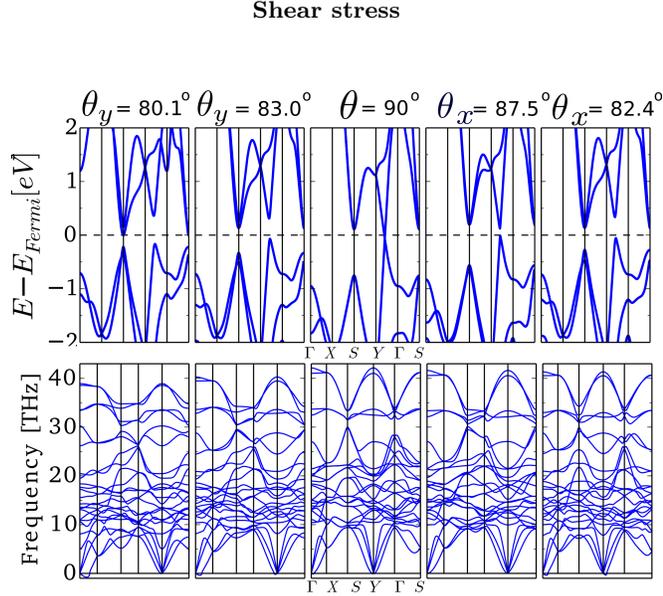}
 \caption{a) Energy band diagrams of 8-{\it Pmmn} borophene with various shearing stress angles. b) Phonon spectra of the corresponding distorted structures.}
 \label{fig4}
\end{figure}

If the formation of the Dirac cones along high-symmetry lines in the Brillouin zone is related to the mirror-reflection symmetry of the material \cite{PhysRevB.93.035401}, it is therefore expected that a distortion of the network from the rectangular to an oblique geometry led likewise to a distortion of the original Dirac cones. Here, we examine the band gap opening obtained upon applying an external shear force to the rectangular cell.

Shear stress consists on a force applied parallel to the structure edges and is equivalent to modifying the 90$^\circ$ angle of the in-plane unit cell vectors of the ground state by adding a finite component to one of the vectors while the other remains unaltered. As a result, the unit cell can be progressively deformed into a slant shape to mimic an external force acting parallel to the x- or y-axis. Such a deformation breaks the internal symmetry of 8-{\it Pmmn} borophene, inducing the distortion of the Dirac cone and the transformation of the material from a metal into an indirect semiconductor. 

In Figure \ref{fig4}, the evolution of the electronic states for different shearing strengths and the phonon spectra of the corresponding distorted structures are in display. For a fixed cell vector in the x-direction, $a_x$, the x-component of the $a_y$ in-plane cell vector is modified gradually, yielding a relative angle $\theta_y$. A reduction of 2$^\circ$ to $\theta_y$=88$^\circ$ is enough to open a band gap (not shown). Additional reduction of  $\theta_y$ to 83.0$^\circ$ and 80.1$^\circ$ remove the degeneracy of some bands and transform distorted 8-{\it Pmmn} borophene into an indirect band gap semiconductor. If the nano-structure is deformed along the perpendicular direction leaving $a_y$ unmodified and increasing the y-component of the $a_x$ vector, the relative angle $\theta_x$ decreases and similar results are obtained. It is worth mentioning that also in the sheared structures the participation of the B$_R$ atoms in the formation of the Dirac cones is negligible.

Although 2D compounds can withstand large deformation stress values, strong reduction of $\theta_x$ or $\theta_y$ can make the nano-structure to become dynamically unstable. The occurrence of an instability is manifested in the phonon spectrum of the distorted nano-structure by negative vibration frequencies. The absence of anomalous softening in the spectra displayed in Figure \ref{fig4} is indicative of the dynamical stability of the skewed structures. Only at the zone center, $\Gamma$, a slight softening is observed for large shear angles, which is related to the restoring force of the material that opposes against deformation.

\subsection{Vacancies}

\begin{figure}[htp]
 \centering
  \includegraphics[width=0.49 \textwidth]{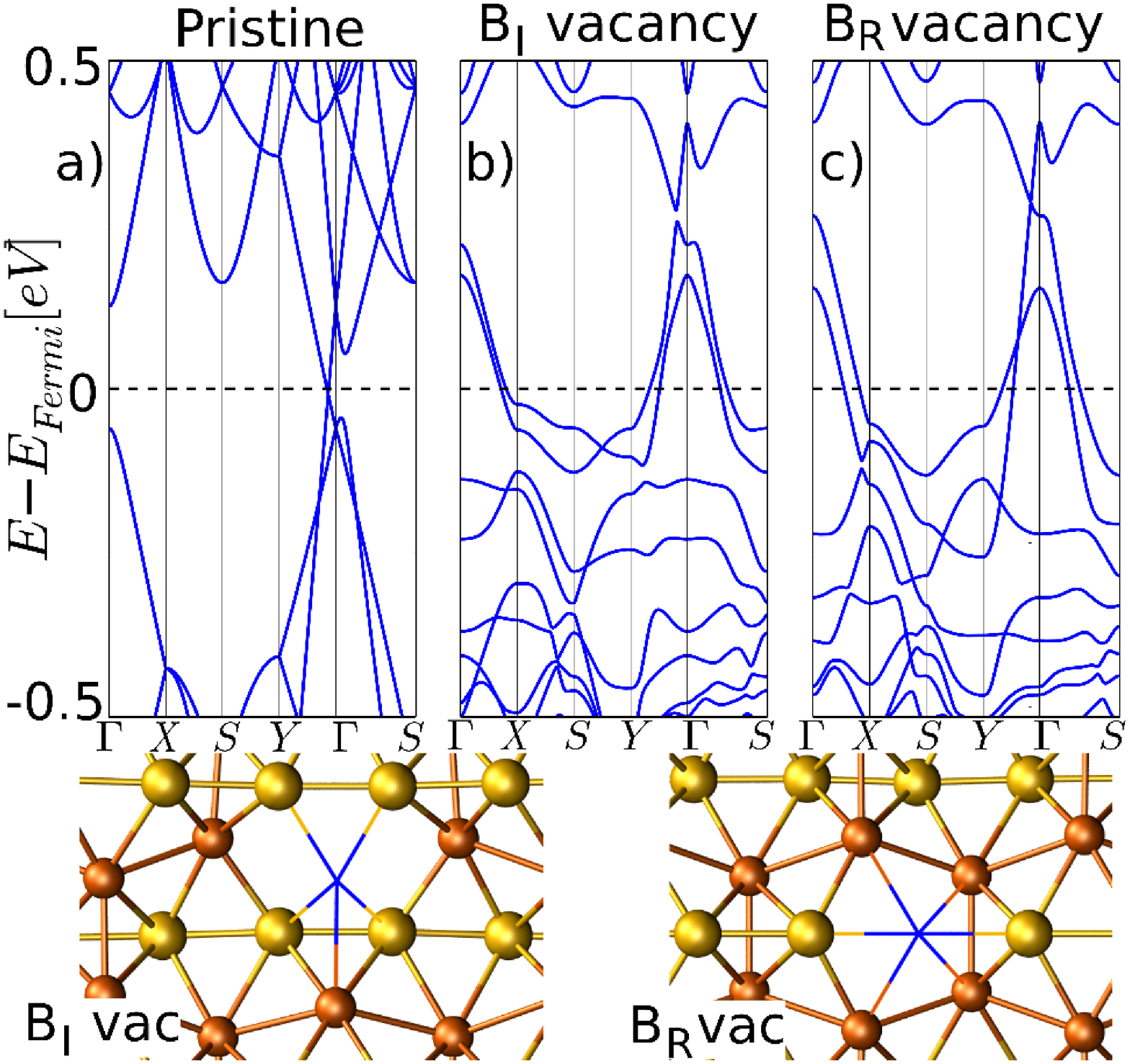}
 \caption{a) Electronic band diagram of a) pristine 5 $\times$ 7 times a 8-{\it Pmmn} borophene unit cell, b) the same supercell with a $B_I$ atom vacant, and c) with a $B_R$ atom vacant. Lower panel show the vacant site and the closest neighboring atoms.}
 \label{fig5}
\end{figure}

The electronic inequivalence of the $B_I$ and $B_R$ atom sublattices has an interesting consequence on the robustness of the electronic properties of 8-{\it Pmmn} borophene against stoichiometric variations.  
Figure \ref{fig5}a shows the first-principles band diagram of a large supercell constructed by repeating 5 $\times$ 7 times a borophene unit cell. Upon removing a B atom from the structure a vacancy is created, which introduces some electronic and geometric distorsion in the structure. One $B_I$ atom missing causes the displacement of the first neighboring atoms to inner positions in the structure, ranging from the vertical 0.25 \AA\ of both $B_R$ and $B_I$ atoms to repulsion of 0.17\AA\ from the vacant site of the $B_I$ atom. Similarly, a neighboring $B_R$ atom is displaced vertically 0.27 \AA\ inwards, while a $B_I$ barely shifts its position by 0.02 \AA\ towards the vacant site when a $B_I$ is missing. 

A vacancy has a different impact on the Dirac cones depending on whether a $B_I$ or a $B_R$ is vacant. In either case the resulting non-magnetic defect shifts down the chemical potential due to the donor character of the vacant site, and the system becomes metallic with two half empty bands crossing the Fermi level. The most striking difference is the appearance of a meV large gap between the bands that originally formed the Dirac cones when the missing atom belongs to the $B_I$ inequivalent sublattice, (Figure \ref{fig5}b). On the contrary, the removal of a $B_R$ atom leaves the reminiscent Dirac cones with no gap in between, (Figure \ref{fig5}c). A $B_R$ vacancy is 0.95 eV energetically more favorable than a $B_I$ vacancy, which adds to the $B_I$ sublattice some protection against structural defects. These results reinforces the autonomy of each type of B atom sublattice in regards to the electronic properties of 8-{\it Pmmn} borophene.

\section{Conclusions}
Many wide-ranging electronic properties of 8-{\it Pmmn} borophene have been detailed to provide an understanding of the potential and flexibility of this material for electronic applications. The nano-structure contains two types of nonequivalent B atoms with opposite effective charges, which identify 8-{\it Pmmn} borophene as of the first single-element based 2D material with a marked ionic character. 
An interesting aspect of this B allotrope is that its metallic behavior is dominated by fragile Dirac cones that can be altered with shear stress. External stress yields a metallic-semiconductor transition in borophene, which turns into a indirect semiconductor for increasing strain. This makes 8-{\it Pmmn} borophene unique in terms of the structural-electronic interplay as compared to other 2D systems such graphene and silicene, since only one of the inequivalent sublattices plays a prominent role in the definition of the material electronic properties.

 \section{Acknowledgments}
We acknowledge DOE BES Glue funding through Grant No. FWP\#70081. We acknowledge the computing resources provided on Blues high-performance computing cluster operated by the Laboratory Computing Resource Center at Argonne National Laboratory. Work at Argonne is supported by DOE-BES under Contract No. DE-AC02-06CH11357.

\end{document}